\documentclass[acmsmall]{acmart}
\usepackage[T1]{fontenc}
\usepackage{amsmath}

\usepackage{amssymb}
\usepackage{enumerate}
\usepackage{bm}
\usepackage{advdate}
\usepackage{datetime}
\usepackage{hyperref}
\usepackage{dsfont}
\usepackage[numbered,framed]{matlab-prettifier}
\usepackage{tkz-euclide}
\usepackage{multirow}
\usepackage{booktabs}
\usepackage[normalem]{ulem}
\usepackage{algorithm}
\usepackage{algorithmic}
\DeclareMathOperator{\vect}{vec}
\usepackage{color}

\AtBeginDocument{%
  \providecommand\BibTeX{{%
    \normalfont B\kern-0.5em{\scshape i\kern-0.25em b}\kern-0.8em\TeX}}}

\setcopyright{acmcopyright}
\copyrightyear{2023}
\acmYear{2023}
\acmDOI{XXXXXXX.XXXXXXX}

\acmConference[Recsys '23]{M}{Sept 18--22,
  2023}{Singapore, SG}
\acmBooktitle{RecSys23',
 Sept. 18--22, 2023, Singapore, SG} 
\acmPrice{15.00}
\acmISBN{978-1-4503-XXXX-X/18/06}

\begin{document}

\title{STAN: Stage-Adaptive Network for Multi-Task Recommendation by Learning \\ User Lifecycle-Based Representation}

\author{Wanda Li}
\authornote{This work was done during Wanda Li’s internship at Shopee.}
\email{lwd20@mails.tsinghua.edu.cn}
\affiliation{%
  \institution{Tsinghua University}
  \country{China}
}

\author{Wenhao Zheng}
\email{zhengwh_nju@foxmail.com}
\author{Xuanji Xiao}\authornote{Corresponding author}
\email{charles.xiao@shopee.com}
\affiliation{%
  \institution{Shopee Company}
  % \streetaddress{1 Th{\o}rv{\"a}ld Circle}
  % \city{Hekla}
  \country{China}}

\author{Suhang Wang}
\affiliation{%
  \institution{Penn State University}
  % \city{Rocquencourt}
  \country{USA}
}

% \affiliation{%

% \affiliation{%

% \affiliation{%

% \affiliation{%

% \affiliation{%


\begin{abstract}
Recommendation systems play a vital role in many online platforms, with their primary objective being to satisfy and retain users. As directly optimizing user retention is challenging, multiple evaluation metrics are often employed. Existing methods generally formulate the optimization of these evaluation metrics as a multi-task learning problem, but often overlook the fact that user preferences for different tasks are personalized and change over time. Identifying and tracking the evolution of user preferences can lead to better user retention.
To address this issue, we introduce the concept of ``user lifecycle,'' consisting of multiple stages characterized by users' varying preferences for different tasks. We propose a novel \textbf{St}age-\textbf{A}daptive \textbf{N}etwork (\textbf{STAN}) framework for modeling user lifecycle stages. STAN first identifies latent user lifecycle stages based on learned user preferences, and then employs the stage representation to enhance multi-task learning performance.
Our experimental results using both public and industrial datasets demonstrate that the proposed model significantly improves multi-task prediction performance compared to state-of-the-art methods, highlighting the importance of considering user lifecycle stages in recommendation systems. Furthermore, online A/B testing reveals that our model outperforms the existing model, achieving a significant improvement of 3.05\% in staytime per user and 0.88\% in CVR. These results indicate that our approach effectively improves the overall efficiency of the multi-task recommendation system.
\end{abstract}

\begin{CCSXML}
<ccs2012>
 <concept>
  <concept_id>10010520.10010553.10010562</concept_id>
  <concept_desc>Computer systems organization~Embedded systems</concept_desc>
  <concept_significance>500</concept_significance>
 </concept>
 <concept>
  <concept_id>10010520.10010575.10010755</concept_id>
  <concept_desc>Computer systems organization~Redundancy</concept_desc>
  <concept_significance>300</concept_significance>
 </concept>
 <concept>
  <concept_id>10010520.10010553.10010554</concept_id>
  <concept_desc>Computer systems organization~Robotics</concept_desc>
  <concept_significance>100</concept_significance>
 </concept>
 <concept>
  <concept_id>10003033.10003083.10003095</concept_id>
  <concept_desc>Networks~Network reliability</concept_desc>
  <concept_significance>100</concept_significance>
 </concept>
</ccs2012>
\end{CCSXML}

\ccsdesc[500]{Information Systems~Collaborative filtering}
\ccsdesc[300]{Machine Learning~Multi-task learning}
\ccsdesc[300]{Machine Learning~Supervised learning by classification}

\keywords{Recommendation Systems, Multi-task Learning, User Lifecycle Modeling}

\maketitle

\section{Introduction}

In recent years, online recommendation systems (RS) have become increasingly popular, assisting users in discovering their preferred items from a vast array of choices on platforms such as e-commerce and social media. The primary objective of RS is to attract, satisfy, and retain users. Researchers have proposed various techniques to achieve these objectives, including multi-task learning methods. 
% However, most prior works overlook the significant influence of the \textit{user lifecycle} on user satisfaction and retention. 
% In this paper, we introduce the concept of user lifecycle and stages, inspired by the common practice in the industry of creating separate models for cold-start and other users.

Due to the high dimensionality of RS \cite{mtl2020cgc}, modeling its objectives is challenging. Many works represent the objectives through multiple directly learnable metrics, such as the likelihood of clicking, forwarding, and staying. Consequently, there has been a growing trend to apply multi-task learning methods to model the various aspects of user interests. Some studies \cite{sigir2020esm2, sigir2018esmm, mtl2020cgc, google2020mose, sigir2022escm} suggest that click-through rate (CTR) and post-click conversion rate (CVR) are the best indicators of user satisfaction, proposing that clicking and purchasing actions are the primary drivers of user retention. Other works \cite{sigir2018essm, ijcai2021dfn, www2022feedrec} consider user feedback (e.g., interactions like forward, comment, stay) as evaluation metrics, assuming that more interactions represent stronger user engagement and aiming to improve all interaction metrics simultaneously.

Nonetheless, these prior works do not fully consider the \textit{user lifecycle} \cite{jure2013lifecycle, kwon2019geolifecycle, li2021deeppick} and its impact on user satisfaction and retention. The \textit{user lifecycle} consists of several stages, each characterized by user preferences towards different tasks. 
These preferences change over time as users evolve, and users may transition between stages with varying probabilities. 

\begin{figure}[]
\vskip -1em
  \includegraphics[width=.9\columnwidth]{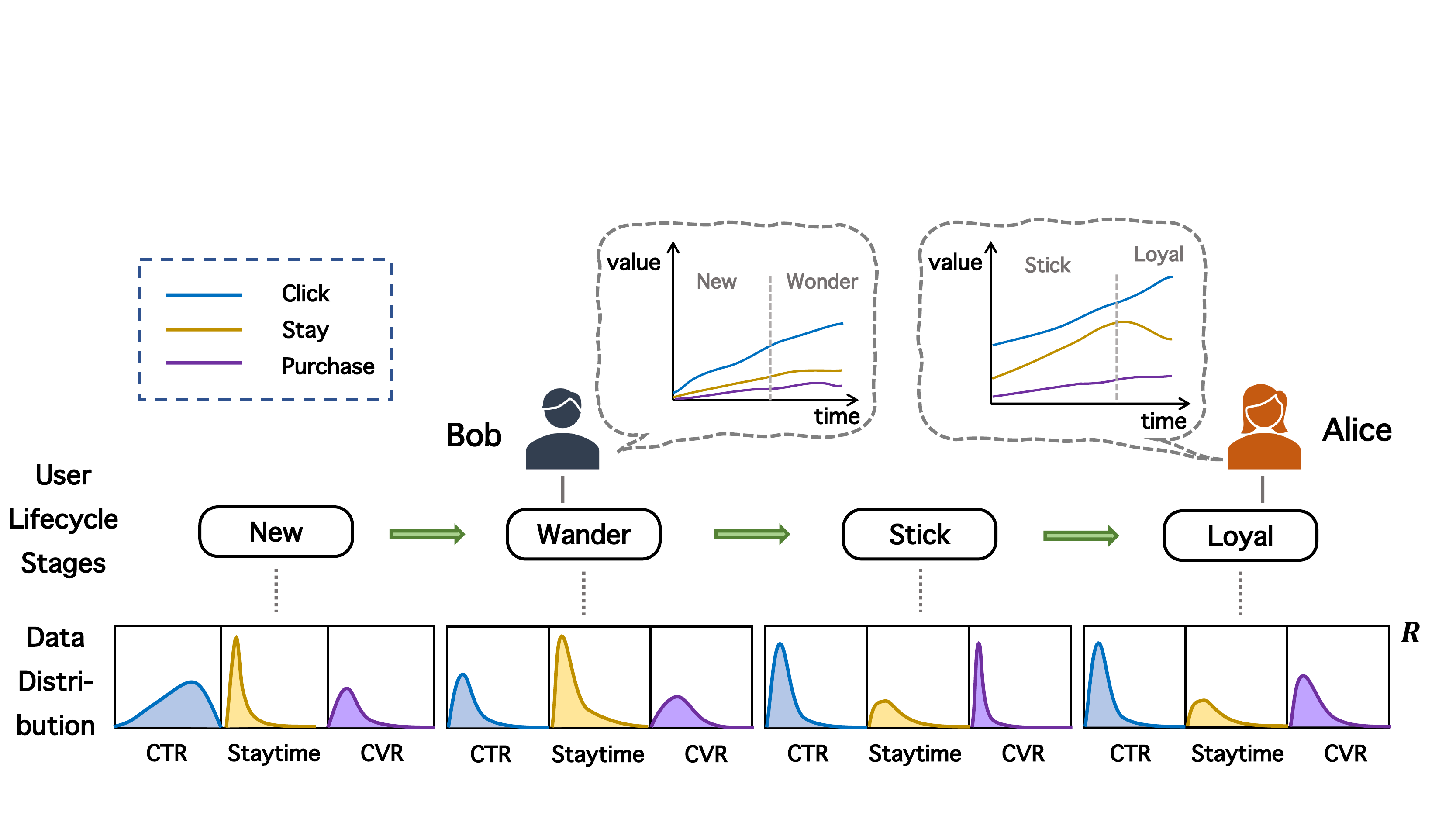}
  \vskip -1em
  \caption{A user's current stage influences his or her preference in real-world recommendation systems.  In the data distribution figures, the x-axis represents the value of CTR/Staytime/CVR, while the y-axis represents the ratio $R$ of users with corresponding values. Discrete stages in the figure are for illustration purposes; in our model, user stages are modeled as continuous vectors.}
  \label{fig1}
\end{figure}

To emphasize the importance of incorporating \textit{user lifecycle} stages into RS, consider the experiences of two users, Bob and Alice, as illustrated in Fig. \ref{fig1}. Typically, a user progresses through a series of stages since registering on the platform. Note that the four discrete stages in Fig. \ref{fig1} are merely examples for visualization purposes, and the actual stages in our model are represented by continuous vectors. 
Bob exhibits wandering behavior, browsing quickly without intending to purchase items. Traditional multi-task RS might try to persuade him to buy by presenting top-selling products. However, this approach could result in his dissatisfaction, causing him to leave without making a purchase. Conversely, Alice has recently transitioned to a more committed stage. She previously preferred browsing to placing orders, but in her current stage, she is more likely to purchase without hesitation if the recommendation suits her preferences. Traditional multi-task RS might persist in recommending items aimed at prolonging her usage time, but these attempts may no longer capture her interest. By taking user lifecycle and stages into account, which can be customized to specific contexts, the recommendation system can more effectively address the diverse needs of users at different stages of their interactions with the platform.

We propose a \textit{user lifecycle} stage-adaptive framework to address these issues. It dynamically adjusts its focus on tasks according to the user's stage, which is modeled by the representation of their preferences. As user behavior may exhibit volatility, it is crucial to account for such instability when modeling preferences. For tasks not aligned with the user's preferences, the user behaviors will be sparse, resulting in actions like clicks and purchases having large randomness and not accurately reflecting users' true preferences. Identifying users' preferences helps the model prioritize reliable targets to learn from, leading to improved performance. By incorporating user stages in the multi-task learning process, the model can focus on users' highest-priority tasks.
Our main contributions can be summarized as follows:

\begin{itemize}
\item To the best of our knowledge, this is the first work to integrate the \textit{user lifecycle} concept into multi-task recommendation systems. By considering the various stages in the \textit{user lifecycle}, we can more effectively capture users' changing interests.
\item We propose a user stage detection network that represents user stages using continuous user preferences, enabling the model to focus dynamically on each user's preferred tasks.
\item We present comprehensive experimental results on both public and industrial datasets to substantiate our claims. Additionally, we applied our method to an online system, achieving significant improvements in online metrics. Visualization results further emphasize the importance of user lifecycle stages in multi-task recommendation systems.
\end{itemize}

The rest of the paper is organized as follows: Sec.~\ref{pda} presents a preliminary analysis on the importance of user stages; Sec.~\ref{method} details our proposed \textbf{St}age-\textbf{A}daptive \textbf{N}etwork (\textbf{STAN}); Sec.~\ref{exp} showcases experiments demonstrating STAN's effectiveness; Sec.~\ref{discussion} discusses the model's expandability and potential; Sec.~\ref{relatedWork} reviews related work, and Sec.~\ref{con} concludes the paper.

\section{Preliminary Analysis}
\label{pda}

In this section, we first conduct a preliminary analysis of real-world datasets to show why considering user stage information is important for RS.
We then formally define the notations and give the problem definition of our work.

\subsection{Insights from Real-world Data}
\label{data}

\begin{figure}[]
  \includegraphics[width=.9\columnwidth]{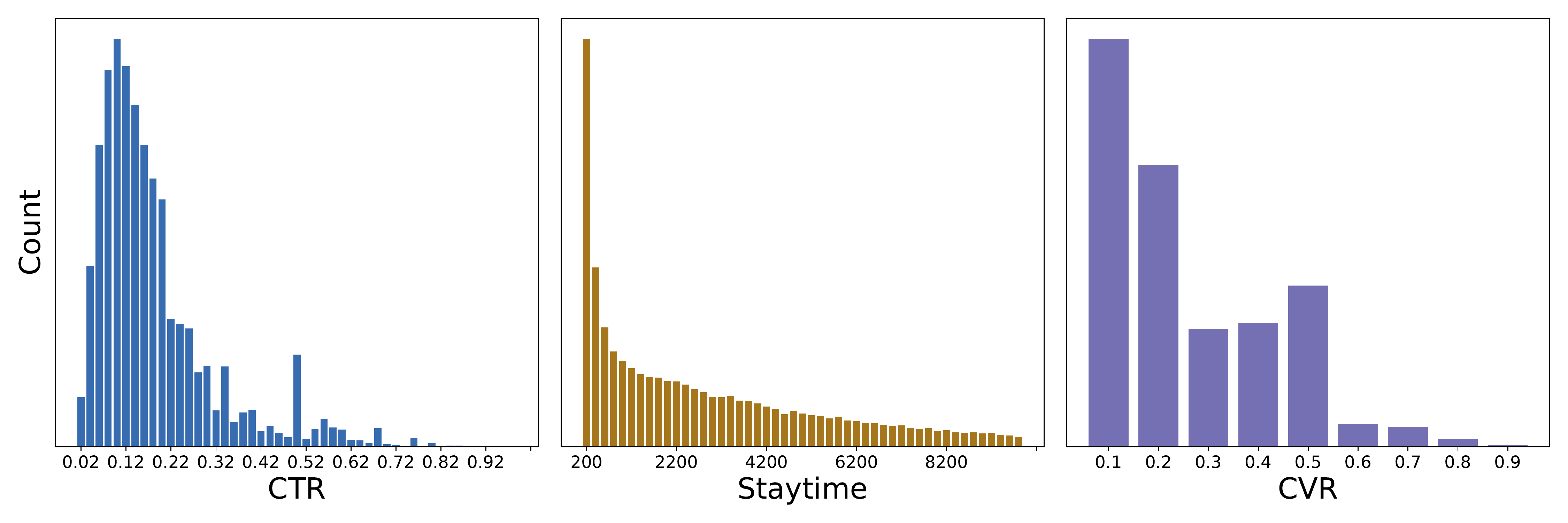}
  \vskip -1em
  \caption{User distribution of three metrics.\protect\footnotemark}
\label{overall}

\end{figure}

In this section, we discuss the e-commerce data used for analysis throughout this paper. We collected one month of user behavior data from an e-commerce platform, which records users' clicks, staytime, and purchase actions. The behavior data is organized by user sessions, each of which is defined as a tuple of actions related to one impressed item.
We randomly selected 50,000 users' actions over three days for data analysis. Fig.~\ref{overall} depicts the user distribution of the three metrics, i.e., CTR, Staytime, and CVR.

\begin{figure}[]
  \includegraphics[width=.9\columnwidth]{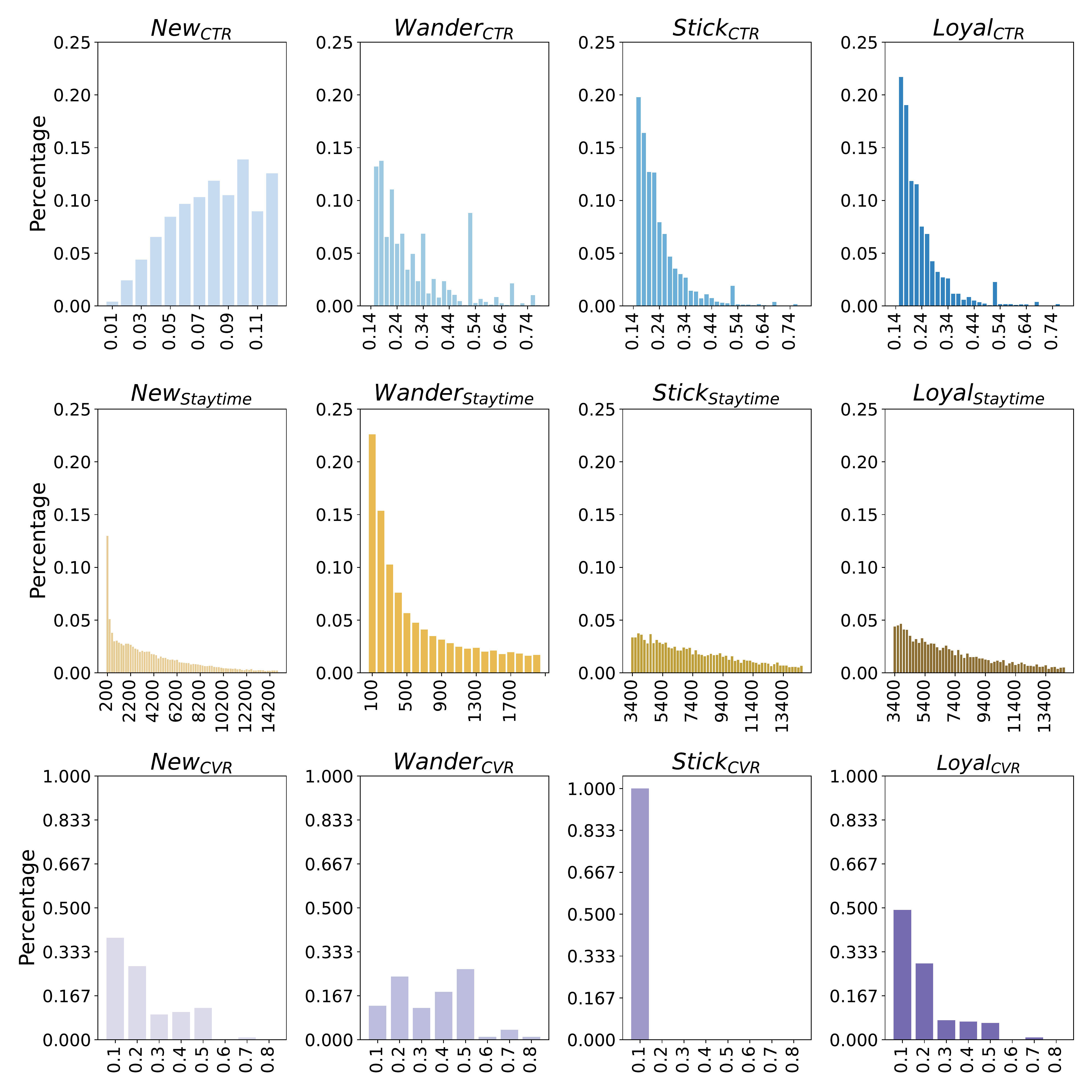}
  \caption{Comparison of user behavior distribution at different stages. Each row shows one metric (i.e., CTR, Staytime, and CVR) and each column shows a stage. The x-axis represents the value of the metric. The y-axis represents the percentage of the x value at the stage. }
  \label{stage}

\end{figure}

Inspired by previous works~\cite{li2021deeppick, kwon2019geolifecycle}, we roughly separate users into four stages by the median of different metrics for a better illustration and explanation. The stages are named \textit{New}, \textit{Wander}, \textit{Stick} and \textit{Loyal}, as shown in Fig.~\ref{stage}. 
Note that one user could not belong to multiple stages. 
Each user at stage \textit{New} are spotted by a low CTR rate, while their staytime and CVR distribution are the most similar to the overall distribution of all users. Their CTR value conforms to a Gaussian-like distribution but in a low-value range. 
Users at stage \textit{Wander} have low staytime length. Their CTR and CVR rates are relatively high, but their staytime stick to a lower range.
Users at stage \textit{Stick} could be quickly found by their relatively low CVR value along with high CTR and staytime, which indicates that they only dwell on the platform but rarely contribute to purchasing.
Users at stage \textit{Loyal} show a custom of steadily clicking, staying, and purchasing on the platform, implying their satisfaction with the recommendation outcomes and the platform.

According to the stage patterns, we notice that users at different stages put their focus on specific things that could be manifested by statistical metrics. If we could discover and suit the user stage in the multi-task recommendation, we are more likely to improve model performance.

\footnotetext{The scale of the y-axis is hidden according to the data sharing policy of the e-commerce platform.}

\subsection{Notations and Problem Definition}

Our preliminary analysis verifies that users at different stages conform to various data distributions. Thus, it would be of great help to consider adaptive stage information while conducting multiple recommendation tasks.

In our cases, we have a dataset $\mathcal{D} = \left\{ \left\{(\mathbf{U}_i, \mathbf{V}_j, \bm{y}_{ij})\right\}_{j=1}^{n_i}\right\}_{i=1}^m$ consisting of $m$ users, each user $i$ exposed to $n_i$ items.  The $\mathbf{U}_i\in\mathbb{R}^{d_1\times d_2}$ represents the user feature matrix, $\mathbf{V}_j\in\mathbb{R}^{d_3\times d_4}$ represents the item feature matrix. Here, 
$d_1$, $d_3$  denote the number of user and item features, respectively, while $d_2$ and $d_4$ represent the dimensions of user and item features, respectively. 
Note that in the preprocessing phase, each user attribute is embedded into a $d_3$-dimensional vector. A similar preprocessing is conducted for items.
The label collection $\bm{y}_{ij}=\left[y_{ij}^1,\dots,y_{ij}^K\right]^{\mathrm{T}} \in \mathbb{R}^K$ includes measurements for $K$ concerned tasks,  where $y_{ij}^k, k=1,\dots,K$ corresponds to  CTR, staytime, and so on. For the overall preference of user $i$, we compute  $\bm{y}_i=\frac{1}{n_i} \sum^{n_i}_{j=1} \left[y_{ij}^1,\dots,y_{ij}^K\right]^{\mathrm{T}} \in \mathbb{R}^K$.
% , where $n_i$ is the number of items the user $i$ have been exposed to.
% The collection of labels $\bm{y}_i=\left[y_i^1,\dots,y_i^K\right] \in \mathbb{R}^K$ concludes the measurements for $K$ concerned tasks, with $y_i^k, k=1,\dots,K$ being CTR, Staytime, etc. 
Note that some user features change over time. In the dataset, the behavior sequence is organized chronologically for each user, but there is no specific order for behavior sequences across different users.

Considering $K$ recommendation prediction tasks, our goal is to develop a framework that can recommend items while taking user stage information into account to enhance the prediction performance for each task.

% \wanda{or: ..learn a stage-detection network $\mathbf{s}(\cdot) \rightarrow \gamma$, where $\mathbf{s}$ is a network we aim to learn and $\gamma$ is the set of task probabilities for the tasks we concerned.}

\section{Proposed Method}
\label{method}
In this section, we propose a framework to extract the latent user stage information for enhancing multi-task recommendation. 

\begin{figure*}[]
  \includegraphics[width=\textwidth]{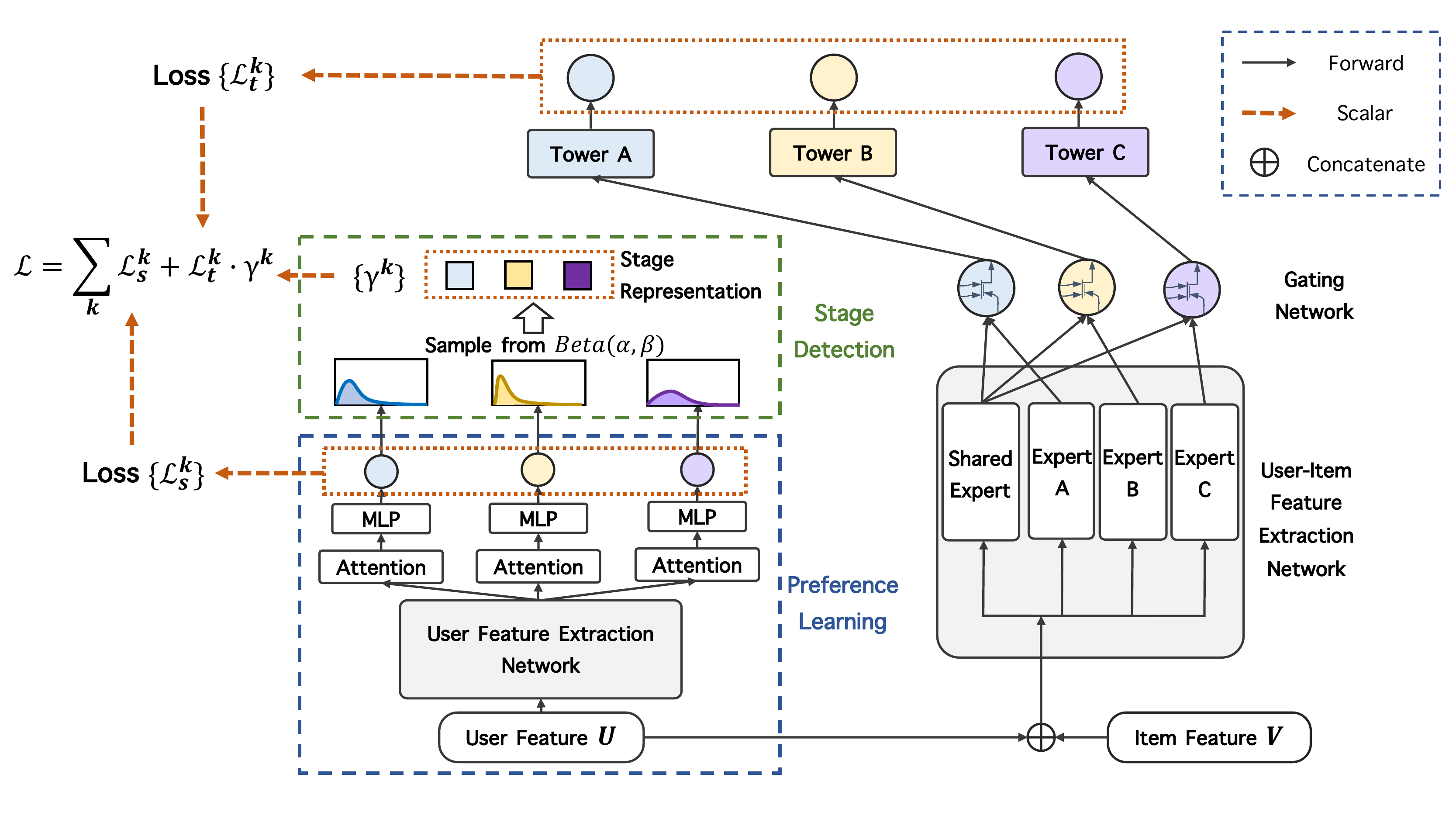}
  \vskip -1em
  \caption{The architecture of the proposed lifecycle-adaptive learning framework. The stage detection network is used to learn the user stage-based task preference for the final prediction.}
  \label{network}
\end{figure*}

To be more specific, the framework is composed of two parts, as shown in Fig. \ref{network}: (i) a multi-task prediction part that learns the representation of the input data for different tasks, and (ii) a latent stage detection part that first grasps user preference.  The preference is then applied to depict the latent user stage. 
Finally, a loss function adaptively adjusts the attention paid to tasks based on the learned user stage. 
The detailed building blocks and functional meanings are illustrated in the following subsections. 

\iffalse
\suhang{For the following paragraphs: 1. start with the ``Feature Extraction Networks''. Actually, it's not feature extraction network, it's already multi-task framework. Discuss it's issue. Then say we need to adptively weight the tasks for each user based on the lifecycle stage}

\suhang{2. After that, you can talk about learning user lifecycle stage followed by the adaptive reweighting.}

\suhang{3. Finally write down the final loss function}
\fi

\subsection{Multi-task Prediction Networks}

In this subsection, we illustrate the components of the multi-task prediction module, which aims to learn the embedding from user and item features by training the backbone network. This network evaluates the difference between the ground truth and the predicted value for each task measurement, as shown on the right-hand side of Fig.~\ref{network}.

In the literature, many multi-task learning models employ one or more shared networks, often referred to as "experts," as the foundation for learning common knowledge among different tasks. While these experts can capture the joint hidden information across tasks, they may suffer from dependencies among tasks and differences in data distribution for various tasks.

To address this issue, we follow the approach of \cite{mtl2020cgc} and limit the usage of shared experts in the first step. By doing so, we aim to alleviate harmful parameter interference, allowing for more robust learning and better performance in capturing the nuances of different tasks within our multi-task prediction module.
%%% TODO: better transition
We use $\vect(\cdot)$ to represent matrix vectorizaion\footnote{For a matrix $\bm{A} = \left[\begin{array}{ll}
a & c \\
b & d
\end{array}\right]$, its vectorization can be written as $\vect(\bm{A}) = \left[\begin{array}{l}
a \\
c \\
b \\
d
\end{array}\right]$.}.
With input $\bm{x}_{ij} = [\vect(\mathbf{U}_i)^{\mathrm{T}},\vect(\mathbf{V}_j)^{\mathrm{T}}]^{\mathrm{T}}\in\mathbb{R}^{d_1d_2 + d_3d_4}, i=1,\dots, m, j=1,\dots, n_i$,
the features are extracted by 
\begin{align}
  w^k_{B}(\bm{x}_{ij}) &= Softmax (\mathbf{W}^k_{B}\bm{x}_{ij}) ,                       \\
  H^k(\bm{x}_{ij}) &= [ Exp^k_{sp} (\bm{x}_{ij}), Exp_{sh} (\bm{x}_{ij})]^{\mathrm{T}},        \\
  g^k_{B} (\bm{x}_{ij}) &= w^k_{B}(\bm{x}_{ij}) \odot H^k(\bm{x}_{ij}),
\end{align}
where $w^k_{B}$ is the weighting function in the backbone network which obtains the weighted vector of task $k$ by a linear layer with the Softmax activation function, $\mathbf{W}^k_{B}$ is the trainable parameter matrix for task $k$ in the backbone, 
$H^k(\bm{x}_{ij})$ is the combination of the task-specific experts $Exp^k_{sp} (\bm{x}_{ij})$ and shared experts  $Exp_{sh} (\bm{x}_{ij})$.
The $g^k_{B}$ is the gating network of task $k$, which acts as the selector to calculate the weighted sum of the input.
The $\odot$ represents the Hadamard (element-wise) product.

We could obtain the prediction value of each task $k$ as:
\begin{align}
  \hat{y}^k_i (\bm{x}_{ij}) &=  f_{T}^k (g^k_{B} (\bm{x}_{ij})),
\end{align}
where $f_{T}^k $ denotes the tower network of task $k, k =1, \dots, K$. %\suhang{add a paragraph explaining a straigtforward loss function of MTL and what's the issue. Then explain what we will do.}

Conventionally, the loss function for multi-task learning can be represented as:
\begin{align}
\mathcal{L} = \sum_k \eta^k \cdot \mathcal{L}^k,
\end{align}
where $\eta^k$ is a hyperparameter and $\mathcal{L}^k$ denotes the task-specific loss function. Typically, $\eta^k$ is determined by heuristic rules. However, using a fixed $\eta^k$ may not be suitable for every user, as their preferences for tasks can vary significantly. Furthermore, even for users whose preferences align with $\eta^k$, the fixed $\eta^k$ might lead to performance degradation when their preferences change over time. To address these issues, we propose the user preference learning module and latent user stage representation module, which dynamically adapt to users' preferences and stages.

\subsection{User Preference Learning}

In this subsection, we introduce the method for extracting user preferences to represent latent user stages. Since there are no explicit criteria to distinguish user stages, we can only infer users' stage information from their behaviors. As the user stage can be described by a set of user preferences for all concerned tasks, we first propose a user preference learning module to represent a user's preference.

The module consists of three building blocks: (i) a user feature extraction network that extracts more representative features from the input user features $\mathbf{U}$, (ii) a task-specific user representation learning unit that generates the corresponding embedding containing the hidden user preference for each task, and (iii) a task measurement prediction unit that utilizes the task-specific embedding to predict the value for each task.

First, the user feature extraction network uses a transformation $f$ that generates the user representation $\mathbf{U}'_i$,
\begin{align}
\mathbf{U}'_i & = f(\mathbf{U}_i),
\end{align}
where $\mathbf{U}'_i \in \mathbb{R}^{d_1 \times d_2}, i=1,\dots,n$. To capture the relationships between different user features, we employ a self-attention unit \cite{nips2017attention} as function $f$. 
The self-attention mechanism excels in modeling long-range dependencies and recognizing important features in the input data, which makes it a suitable choice for extracting user preferences. 
Specifically, let $\mathbf{W}_Q\mathbf{U}_i$, $\mathbf{W}_K\mathbf{U}_i$, and $\mathbf{W}_V\mathbf{U}_i$ be the query, key, and value matrices, where $\mathbf{W}_Q, \mathbf{W}_K$, and $\mathbf{W}_V$ are all $d_1$ by $d_1$ weight matrices. Then, the self-attention mechanism can be written as
\begin{align}
f(\mathbf{U}_i) = Softmax\left(\frac{1}{\sqrt{d_1}}\mathbf{W}_Q\mathbf{U}_i(\mathbf{W}_K\mathbf{U}_i)^{\mathrm{T}} \right)\mathbf{W}_V\mathbf{U}_i,
\end{align}
where $Softmax(\cdot)$ is the softmax function\footnote{For vector $\bm{z}=[z_1,\dots, z_d]^{\mathrm{T}}\in\mathbb{R}^d$, the $j$-th entry of the softmax function is $Softmax(\bm{z})_j= \dfrac{e^{z_j}}{\sum_{i=1}^d e^{z_i}}$.}.

Next, another attention unit generates a task-specific user representation according to the importance of different features, inspired by \cite{aaai2021attention}. 
The attention mechanism effectively captures and emphasizes the importance of various features within the user representation. It allows the model to focus on the most relevant features for each specific task, ultimately leading to more accurate and tailored task-specific user representations.
For task $k$, the corresponding embeddings $\mathbf{s}_i^k, k = 1, \dots, K$, are computed as follows,
% \begin{align}
%   \mathbf{s}_i^k & = \sum_{j=1}^{d_1}  \mathbf{U}'_i(j,:)^{\mathrm{T}} \odot Softmax(\mathbf{W}^k \cdot \mathbf{U}'_i(j,:)^{\mathrm{T}}),\quad k = 1, \dots, K,
% \end{align}
\begin{align}
  \mathbf{s}_i^k & = \mathbf{U}'_i \odot Softmax(\mathbf{W}^k \cdot \mathbf{U}'_i),\quad k = 1, \dots, K,
\end{align}
where the weight matrix $\mathbf{W}^k\in\mathbb{R}^{d_2\times d_2}$ for task $k$ is learned during training.

Finally, the task measurement prediction unit takes in the task-specific user embedding to produce predicted values $\tilde{y}^k_i$ for each task. The prediction unit learns to represent the general user preference irrespective of individual items by training with user interactions across all items the user has encountered.
%\suhang{It makes sense if you call it preference towards task $k$. However, it doesn't make sure if you use $y_i^k$ as the groundtruth as $y_i^k$ is click/purchase/stay for a specific item; while the preference towards task $k$ is a general preference.}

Single-layer feed-forward networks $\mathrm{MLP}^{k}(\cdot), k=1, \dots, K$ are used to predict the probability for the user to conduct the corresponding actions. 
Based on the learned embedding, the output user preference is defined as: 
\begin{align}
  \tilde{y}^k_i = Sigmoid(\mathrm{MLP}^{k}(\mathbf{s}_i^k)), i=1, \dots, n; k=1,\dots, K.
\end{align}

% TODO: it's not a label -> pseudo label -> don't use directly as it's uncertain. use network to learn more information. hard so former works not done.
% why not use l for \gamma (in practice, that is )
% During the training process, the label of $\tilde{y}_i^k$ can be defined as follows:

To account for the volatility in user behavior, we create a pseudo-label by considering the average behavior value over time for $\tilde{y}_i^k$. This approach helps the network learn more information while maintaining greater stability:
\begin{align}
\label{eq_l}
    l^k_i=\frac{1}{|\mathcal{D}_i|} \sum_{y_j^k \in \mathcal{D}_i} y_j^k,
\end{align}
where $\mathcal{D}_i=\bigcup_{j=1}^{n_{i'}} \{(\mathbf{U}_{i'},\mathbf{V}_j,\bm{y}_{i'})| i' < i\}$ is the subset of $\mathcal{D}$ containing only the first $i$ instances. This method incorporates the user's preferences from the past few days, ensuring a more accurate representation of their recent interests.

Subsequently, the loss can be defined as:

\begin{align}
    \label{loss_l}
    \mathcal{L}^k_s(\tilde{y}^k, l^k) & = \sum_{i=1}^n (l_i^k -  \tilde{y}_i^k)^2.
\end{align}

One advantage of our user preference learning module is that it can capture user preferences without explicit supervision.
Moreover, the learned preference representation could be adjusted dynamically as user behaviors change. Thus, the preference modeling process is of great adaptation to user preferences.
% which is the traditional method that determining the user stage based on particular man-made rules failed to \textit{adaptive} to. 
% As different online platforms may have few in common, learning the user stage adaptively could enhance the transferability of the RS algorithm from one dataset to another.

\subsection{Latent User Stage Detection}
\label{stage_ad}

In this subsection, we develop a latent user stage representation module, drawing inspiration from prior successful RS applications \cite{beta2014coldstart, beta2011click}. We regard the output of the user preference module, $\tilde{y}^k, k=1,\dots, K$, as the user's inclination towards task $k$. However, when the number of training samples for a user is limited, the predicted $\tilde{y}^k$ may be less informative. Therefore, we introduce the latent user stage representation module to adjust $\tilde{y}^k$ and generate a reliable preference, $\gamma^k$, for tasks during subsequent training.

More specifically, based on the initial data analysis in Sec.~\ref{pda}, the value $\tilde{y}^k$ is assumed to follow a Beta distribution, $Beta(\alpha, \beta)$, i.e.,

\begin{align}
\tilde{y}^k &\sim Beta(\alpha^k, \beta^k),
\end{align}

where $\alpha^k$ represents the number of trials in which user $u$ performs the concerned action in task $k$, and $\beta^k$ represents the number of trials in which user $u$ does not perform the action. For example, when considering the task of predicting CTR, $\alpha$ corresponds to the number of clicks, while $\beta$ represents the number of times the user did not click on the item.

The parameters $\alpha^k$ and $\beta^k$ can be learned during training.
As shown in Algorithm~\ref{alg1}, the learned user preference toward different tasks will be more reliable as more samples are fed into the latent user stage representation module.
% Moreover, the calculation of parameters $\alpha^k$ and $\beta^k$ put emphasis on $\tilde{y}^k$ with the increasing occurrence of the user. 
% Thus, the model is preferable to the near-time behavior. 
With the refined behavior preference distribution $Beta(\alpha^k, \beta^k)$, the output $\tilde{y}^k$ 
 will be improved owing to a more robust and reliable preference $\gamma^k$.

Then, the preference measurement $\gamma^k$ is used to control loss importance corresponding to user attention at the current stage.
For the task in which the user is less interested, the corresponding loss will be given less attention in the back-propagation process. It could reduce the risk of negative transfer or seesaw phenomenon~\cite{mtl2020cgc} in the extraction backbone network $f^{B}$. Thus, the representation $f^{B}(x)$ could be learned across tasks with a concentration on the valuable information about the user for the tasks.

\begin{algorithm}
	\renewcommand{\algorithmicrequire}{\textbf{Input:}}
	\renewcommand{\algorithmicensure}{\textbf{Output:}}
        \newcommand{\1}[1]{\mathds{1}\left[#1\right]}
    \caption{Process of updating $Beta(\alpha^k, \beta^k)$ and sampling $\gamma^k$ }
    \label{alg1}
    \begin{algorithmic}[1]
    \REQUIRE ~~     \\
    $\left\{ (\mathbf{U}_i, \bm{y}_i)\right\}_{i=1}^n$. Note that a higher index indicates a time closer to the present.
    \ENSURE ~~  \\
     $\gamma^k$: The series of sampling results, the latent stage representation for task $k$.
    \FORALL{$k$ in 1,\dots, K,}
        \FORALL{$u$ in 1,\dots, m,}
            \STATE Initialize $\alpha_u^k = \beta_u^k =1$. $\gamma_u^k = c_u =0$,
            \FORALL{$i$ in 1,\dots, n,}
                \IF{${\rm user\ id\ of\ } U_i = u$}
                    \STATE $c_u \leftarrow c_u + 1$,
                    \STATE $\alpha_u^k \leftarrow \alpha_u^k + \tilde{y}_i^k\cdot c_u$,
                    \STATE $\beta_u^k \leftarrow \beta_u^k + (1-\tilde{y}_i^k)\cdot c_u$.
                \ENDIF
            \ENDFOR
            \STATE Draw $\gamma_u^k$ from $Beta(\alpha_u^k, \beta_u^k)$.
        \ENDFOR
    \ENDFOR
    \end{algorithmic}
\end{algorithm}

% different gate serves for different stage, which has different label (for the same input Emb(x).

\subsection{Loss Function of STAN}
There are two parts of losses in our framework: (i) loss $\mathcal{L}^k_t$ for multi-task prediction network, and (ii) loss $\mathcal{L}^k_s$ for stage detection network. Both of them are with respect to task $k, k = 1, \dots, K$. Thus, we give the overall objective to minimize as a linear combination of the losses with task-specific preference $\gamma^k$:
\begin{align}
  \mathcal{L}(\hat{y}^k, \tilde{y}^k, y^k, l^k) = \sum_{k=1}^K  \left( \gamma^k \cdot \mathcal{L}^k_t(\hat{y}^k, y^k) + \mathcal{L}^k_s(\tilde{y}^k, l^k) \right),
\end{align}
where $\mathcal{L}^k_s(\tilde{y}^k, l^k)$ is defined in Eq.~\ref{loss_l}.
Let $y^k_i \in \{0, 1\}$ be the label in dataset, the $\mathcal{L}^k_t(\hat{y}^k, y^k)$ is defined as:
\begin{align}
  \mathcal{L}^k_t(\hat{y}^k, y^k) = \sum_{i=1}^n \left( y_i^k \log \hat{y}_i^k + (1 - y_i^k)\log (1 - \hat{y}_i^k)\right).
\end{align}

Note that the losses for user stage detection and multi-task prediction are optimized simultaneously during training. In this way, the assessment of the user lifecycle stage can be optimized as the loss $\mathcal{L}$ decreases. Improved stage representation can aid in enhancing the multi-task recommendation.

\section{Experiments}
\label{exp}
In this section, we conduct comprehensive offline and online experiments on both large-scale recommendation systems and public benchmark datasets to evaluate the effectiveness of our proposed framework. Our experiments aim to address the following research questions:

\textbf{RQ1:} What is the performance of our proposed method compared with other state-of-art methods? 

\textbf{RQ2:} What effect does detecting the user's stage have on prediction results?

\textbf{RQ3:} Is the proposed framework able to effectively detect user lifecycle stages?

\subsection{Experimental Settings}

In this part, we provide an overview of the dataset descriptions, the baseline methods used for comparison, hyperparameters, and evaluation metrics employed in our experiments.

\subsubsection{Datasets}

\begin{table}[]
\caption{Dataset description.}
  \resizebox{.5\columnwidth}{!}{
      \begin{tabular}{@{}ccccc@{}}
      \toprule
      Dataset      & \# User &  \# Train &  \# Valid &  \# Test \\ \midrule
      Wechat-Video & 0.02M   & 6.71M                            & 0.61M                            & 0.43M                           \\
      Industrial   & 75.42M  & 612.88M                          & 82.47M                           & 407.95M                         \\ \bottomrule
      \end{tabular}
  }
  \label{data_description}
\end{table}

\begin{itemize}
\item Public dataset\footnote{https://algo.weixin.qq.com/2021/problem-description}: The Wechat-Video dataset is a publicly available dataset containing 7.3 million user interaction samples from the Wechat Channels' Recommendation System, involving a total of 20,000 users. Since no existing dataset provides a comprehensive set of tasks, including CVR, staytime, and CVR prediction, we utilize the most common user interactions in the Wechat-Video dataset, such as like, click avatar, and forward. 
\item Industrial dataset: This dataset was collected from an e-commerce platform over a month in 2022. It consists of offline logs from one scenario in the livestreaming recommendation system and is chronologically divided into training, validation, and test sets. Since the staytime of users is a continuous value, we apply equal-frequency binning to the staytime in the dataset for convenience. Further analysis of the dataset can be found in Sec.~\ref{data}.
\end{itemize}

\subsubsection{Baseline Methods}

We compare our methods with the following competitive baselines:

\begin{itemize}
\item $\mathbf{MLP_{Single}}$ \cite{mlp2016deep}: A single-task learning model using a basic MLP (multi-layer perceptron) for each task.
\item $\mathbf{MLP_{Shared}}$ \cite{sharedbottom1997}: A shared-bottom model that shares the bottom layer in multi-task learning, implemented by MLP.
\item \textbf{MMOE} \cite{kdd2018mmoe}: Uses a shared \textit{Expert} module to extract underlying feature embeddings and applies different gates for each task to obtain varying fusing weights in multi-task learning.
\item $\mathbf{PLE_{vanilla}}$ \cite{mtl2020cgc}: PLE (Progressive Layered Extraction) explicitly separates task-common and task-specific parameters to avoid parameter conflicts in multi-task learning using Customized Gate Control (CGC) layers. A PLE model consists of multiple CGC layers.
\item \textbf{AITM} \cite{kdd2021aitm}: AITM (Adaptive Information Transfer Multi-task) models the sequential dependence among audience multi-step conversions using an information transfer module, focusing on different conversion stages of various audiences.
\item $\mathbf{PLE_{Stage}}$: Enhances the vanilla PLE structure by adding pre-knowledge of stage information. Stage information is combined with existing features. PLE is a widely applied baseline in the multi-task recommendation field, and we use it to evaluate the effect of incorporating stage information.
\end{itemize}

To examine the impact of adopting the Beta distribution, we conduct experiments on STAN without the latent stage representation module in Sec.~\ref{stage_ad}, named \textbf{STAN w/o Beta}.

\subsubsection{Hyper-Parameter}

For a fair comparison, we search for optimal parameters on the validation data and evaluate these models on the test data. To ensure a level playing field, we constrain the maximum model size for all methods by setting the same upper bound for the number of hidden units per layer at 1024. For computational efficiency, we assign an embedding dimension of 128 to all methods. We employ ReLU~\cite{relu2011deep} as the activation function for all models. During training, we set the batch size to 2048. The Adam optimizer~\cite{kingma2014adam} is used with settings $\beta_1$ = 0.9, $\beta_2$ = 0.999, and $\epsilon = 1 \times 10^{-6}$. We set the learning rate at 0.001.

For the user feature learning function $f$, we adopt a deep neural network (DNN) structure due to its promising ability to extract hidden information from embeddings.  Our approach and all baseline methods are implemented using TensorFlow\footnote{https://www.tensorflow.org/}.

\subsubsection{Evaluation Metrics}

Specifically, we aim to evaluate the proposed work on two fronts: prediction and ranking.
For offline experiments, existing works primarily use AUC (Area Under ROC) as the main ranking metric to gauge model performance. However, AUC only evaluates the average ranking performance of the model at all thresholds, disregarding the true user interest for each recommendation feed.
Therefore, we apply Normalized Discounted Cumulative Gain (NDCG), which is suitable for evaluating whether users are generally interested in the top recommended items.
Due to the specificity of the industrial dataset, the impression history of users is relatively short. In many cases, only a minimal number of impressions are collected.
As a result, we use NDCG@1 as the evaluation metric for the industrial dataset.
For the public dataset, we use NDCG@5 instead.

For comparison, we follow \cite{kdd2018din} to introduce RelaImpr metric to measure the relative improvement of a measured model over the base model. For a random guesser, the value of AUC is 0.5. Hence RelaImpr for AUC is defined as:
\begin{align}
  RelaImpr = \left( \frac{\rm AUC(measured \ model)-0.5}{\rm AUC(base \ model)-0.5} -1 \right) \times 100\%.
\end{align}

For NDCG@$k'$ ($k' \in \left\{ 1,5 \right\}$ in our experiments), the RelaImpr is defined as:
\begin{align}
  RelaImpr = \left( \frac{{\rm NDCG}@k'({\rm measured \ model})}{{\rm NDCG}@k'({\rm base \ model})} -1 \right) \times 100\%.
\end{align}
In our experiments, we choose $MLP_{Single}$ as the base model.

\useunder{\uline}{\ul}{}
\begin{table*}[]
\caption{Comparison with State-of-the-arts in the public dataset.  ``*'' marks the methods that improve the best baselines, which are underlined, significantly at p-value < 0.01 over paired samples t-test.}
\label{pub}
\renewcommand\arraystretch{1.1}
\resizebox{\textwidth}{!}{
    \begin{tabular}{@{}cccccccccc@{}}
\toprule
\multirow{2}{*}{\textbf{Label}} & \multirow{2}{*}{\textbf{Metric}} & \multicolumn{8}{c}{\textbf{Method}}                                                                                                                                                                                                                    \\ \cmidrule(l){3-10} 
                                &                                  & \multicolumn{1}{c|}{$\mathbf{MLP_{Single}}$}   & $\mathbf{MLP_{Shared}}$ & \textbf{MMOE}     & $\mathbf{PLE_{vanilla}}$        & \multicolumn{1}{c|}{\textbf{AITM}}       & \multicolumn{1}{c|}{$\mathbf{PLE_{stage}}$} & \textbf{STAN w/o Beta} & \textbf{STAN}               \\ \midrule
\multirow{4}{*}{Like}           & \multicolumn{1}{c|}{AUC}         & \multicolumn{1}{c|}{0.8233$\pm$0.0003} & 0.8241$\pm$0.0007      & 0.8260$\pm$0.0019 & {\ul 0.8317$\pm$0.0011} & \multicolumn{1}{c|}{0.8312$\pm$0.0009}   & \multicolumn{1}{c|}{0.8313$\pm$0.0011}   & 0.8413±0.0023      & \textbf{0.8419$\pm$0.0006*} \\
                                & \multicolumn{1}{c|}{RelaImpr}    & \multicolumn{1}{c|}{-}                 & 0.24\%                 & 0.83\%            & 2.59\%                  & \multicolumn{1}{c|}{2.44\%}              & \multicolumn{1}{c|}{2.47\%}              & 5.58\%             & 5.76\%                      \\
                                & \multicolumn{1}{c|}{NDCG@5}      & \multicolumn{1}{c|}{0.5652±0.0037}     & 0.5793±0.0062          & 0.6042±0.0055     & 0.6049±0.0065           & \multicolumn{1}{c|}{{\ul 0.6437±0.0032}} & \multicolumn{1}{c|}{0.6418±0.0071}       & 0.6645±0.0083      & \textbf{0.6737±0.0053*}     \\
                                & \multicolumn{1}{c|}{RelaImpr}    & \multicolumn{1}{c|}{-}                 & 1.42\%                 & 3.90\%            & 3.98\%                  & \multicolumn{1}{c|}{7.86\%}              & \multicolumn{1}{c|}{7.67\%}              & 9.93\%             & 10.85\%                     \\ \midrule
\multirow{4}{*}{Click\_avatar}  & \multicolumn{1}{c|}{AUC}         & \multicolumn{1}{c|}{0.8230±0.0022}     & 0.8279±0.0020          & 0.8286±0.0018     & 0.8317±0.0015           & \multicolumn{1}{c|}{0.8442±0.0019}       & \multicolumn{1}{c|}{{\ul 0.8471±0.0023}} & 0.8496±0.0074      & \textbf{0.8530±0.0021*}     \\
                                & \multicolumn{1}{c|}{RelaImpr}    & \multicolumn{1}{c|}{-}                 & 1.52\%                 & 1.73\%            & 2.69\%                  & \multicolumn{1}{c|}{6.57\%}              & \multicolumn{1}{c|}{7.47\%}              & 8.22\%             & 9.30\%                      \\
                                & \multicolumn{1}{c|}{NDCG@5}      & \multicolumn{1}{c|}{0.2129±0.0051}     & 0.2261±0.0036          & 0.2295±0.0047     & {\ul0.2371±0.0053 }          & \multicolumn{1}{c|}{0.1993±0.0084}       & \multicolumn{1}{c|}{ 0.2371±0.0060} & 0.2474±0.0089      & \textbf{0.2499±0.0044*}     \\
                                & \multicolumn{1}{c|}{RelaImpr}    & \multicolumn{1}{c|}{-}                 & 1.32\%                 & 1.67\%            & 2.42\%                  & \multicolumn{1}{c|}{-1.36\%}             & \multicolumn{1}{c|}{2.42\%}              & 3.45\%             & 3.70\%                      \\ \midrule
\multirow{4}{*}{forward}        & \multicolumn{1}{c|}{AUC}         & \multicolumn{1}{c|}{0.8692±0.0005}     & 0.8751±0.0011          & 0.8803±0.0010     & 0.8805±0.0003           & \multicolumn{1}{c|}{0.86978±0.0019}      & \multicolumn{1}{c|}{{\ul 0.8837±0.0006}} & 0.8846±0.0040      & \textbf{0.8856±0.0009}      \\
                                & \multicolumn{1}{c|}{RelaImpr}    & \multicolumn{1}{c|}{-}                 & 1.58\%                 & 2.99\%            & 3.03\%                  & \multicolumn{1}{c|}{0.14\%}              & \multicolumn{1}{c|}{3.90\%}              & 4.25\%             & 4.42\%                      \\
                                & \multicolumn{1}{c|}{NDCG@5}      & \multicolumn{1}{c|}{0.1317±0.0011}     & {\ul 0.1536±0.0017}    & 0.1515±0.0014     & 0.1405±0.0006           & \multicolumn{1}{c|}{0.1392±0.0015}       & \multicolumn{1}{c|}{0.1483±0.0016}       & 0.1570±0.0039      & \textbf{0.1581±0.0013*}     \\
                                & \multicolumn{1}{c|}{RelaImpr}    & \multicolumn{1}{c|}{-}                 & 2.20\%                 & 1.99\%            & 0.88\%                  & \multicolumn{1}{c|}{0.75\%}              & \multicolumn{1}{c|}{1.67\%}              & 2.53\%             & 2.64\%                      \\ \bottomrule
\end{tabular}
}
  \end{table*}

\begin{table*}[]
\centering
        \caption{Comparison with State-of-the-arts in the industrial dataset. ``*'' marks the methods that improve the best baselines, which are underlined, significantly at p-value < 0.01 over paired samples t-test.}
        \label{ind}
\renewcommand\arraystretch{1.1}
\resizebox{\textwidth}{!}{
    \begin{tabular}{@{}cccccccccc@{}}
\toprule
\multirow{2}{*}{\textbf{Label}} & \multirow{2}{*}{\textbf{Metric}} & \multicolumn{8}{c}{\textbf{Method}}                                                                                                                                                                                                                    \\ \cmidrule(l){3-10} 
                                &                                  & \multicolumn{1}{c|}{$\mathbf{MLP_{Single}}$}   & $\mathbf{MLP_{Shared}}$ & \textbf{MMOE}     & $\mathbf{PLE_{vanilla}}$        & \multicolumn{1}{c|}{\textbf{AITM}}       & \multicolumn{1}{c|}{$\mathbf{PLE_{stage}}$}     & \textbf{STAN w/o Beta} & \textbf{STAN}               \\ \midrule
\multirow{4}{*}{CTR}            & \multicolumn{1}{c|}{AUC}         & \multicolumn{1}{c|}{0.7891$\pm$0.0015} & 0.7938$\pm$0.0019      & 0.8066$\pm$0.0016 & 0.8056$\pm$0.0014   & \multicolumn{1}{c|}{0.8071$\pm$0.0011}   & \multicolumn{1}{c|}{{\ul 0.8089$\pm$0.0013}} & 0.8140±0.0038      & \textbf{0.8141$\pm$0.0010*} \\
                                & \multicolumn{1}{c|}{RelaImpr}    & \multicolumn{1}{c|}{-}                 & 1.63\%                 & 6.05\%            & 5.71\%              & \multicolumn{1}{c|}{6.23\%}              & \multicolumn{1}{c|}{6.85\%}                  & 8.61\%             & 8.65\%                      \\
                                & \multicolumn{1}{c|}{NDCG@1}      & \multicolumn{1}{c|}{0.5727±0.0132}     & 0.5945±0.0140          & 0.5895±0.0129     & 0.5825±0.0122       & \multicolumn{1}{c|}{{\ul 0.6440±0.0137}} & \multicolumn{1}{c|}{0.6421±0.0124}           & 0.6739±0.0159      & \textbf{0.6737±0.0128*}     \\
                                & \multicolumn{1}{c|}{RelaImpr}    & \multicolumn{1}{c|}{-}                 & 2.19\%                 & 1.68\%            & 0.98\%              & \multicolumn{1}{c|}{7.12\%}              & \multicolumn{1}{c|}{6.94\%}                  & 10.12\%            & 10.10\%                     \\ \midrule
\multirow{4}{*}{Staytime}       & \multicolumn{1}{c|}{AUC}         & \multicolumn{1}{c|}{0.6635±0.0021}     & 0.6721±0.0041          & 0.6770±0.0024     & 0.6835±0.0025       & \multicolumn{1}{c|}{0.6801±0.0023}       & \multicolumn{1}{c|}{{\ul 0.6879±0.0028}}     & 0.6925±0.0043      & \textbf{0.6937±0.0025*}     \\
                                & \multicolumn{1}{c|}{RelaImpr}    & \multicolumn{1}{c|}{-}                 & 5.27\%                 & 8.26\%            & 12.29\%             & \multicolumn{1}{c|}{10.15\%}             & \multicolumn{1}{c|}{14.92\%}                 & 17.74\%            & 18.47\%                     \\
                                & \multicolumn{1}{c|}{NDCG@1}      & \multicolumn{1}{c|}{0.8029±0.0018}     & 0.8186±0.0010          & 0.8270±0.0039     & 0.8170±0.0023       & \multicolumn{1}{c|}{0.7993±0.0021}       & \multicolumn{1}{c|}{{\ul 0.8273±0.0016}}     & 0.8381±0.0029      & \textbf{0.8399±0.0028*}     \\
                                & \multicolumn{1}{c|}{RelaImpr}    & \multicolumn{1}{c|}{-}                 & 1.58\%                 & 2.41\%            & 1.41\%              & \multicolumn{1}{c|}{-0.36\%}             & \multicolumn{1}{c|}{2.44\%}                  & 3.52\%             & 3.70\%                      \\ \midrule
\multirow{4}{*}{CVR}            & \multicolumn{1}{c|}{AUC}         & \multicolumn{1}{l|}{0.7934±0.0057}     & 0.7998±0.0063          & 0.8090±0.0069     & {\ul 0.8217±0.0073} & \multicolumn{1}{c|}{0.8131±0.0081}       & \multicolumn{1}{c|}{0.8192±0.0052}           & 0.8267±0.0090      & \textbf{0.8304±0.0075}      \\
                                & \multicolumn{1}{c|}{RelaImpr}    & \multicolumn{1}{c|}{-}                 & 2.19\%                 & 5.32\%            & 9.65\%              & \multicolumn{1}{c|}{6.71\%}              & \multicolumn{1}{c|}{8.79\%}                  & 11.35\%            & 12.62\%                     \\
                                & \multicolumn{1}{c|}{NDCG@1}      & \multicolumn{1}{c|}{0.5617±0.0023}     & 0.6125±0.0035          & 0.6198±0.0025     & {\ul 0.6284±0.0027} & \multicolumn{1}{c|}{0.6026±0.0032}       & \multicolumn{1}{c|}{0.6250±0.0028}           & 0.6578±0.0041      & \textbf{0.6601±0.0021*}     \\
                                & \multicolumn{1}{c|}{RelaImpr}    & \multicolumn{1}{c|}{-}                 & 5.08\%                 & 5.81\%            & 6.67\%              & \multicolumn{1}{c|}{4.09\%}              & \multicolumn{1}{c|}{6.33\%}                  & 9.61\%             & 9.84\%                      \\ \bottomrule
\end{tabular}
}
\end{table*}

\subsection{Performance Evaluation}

To answer \textbf{RQ1}, we conduct experiments to compare the proposed model with the baseline methods.
For the public dataset, we consider common actions, including like, click avatar (of the content creator), and forward.
For the industrial dataset, we focus on classical RS prediction tasks: CTR, staytime, and CVR.
The corresponding results are shown in Table~\ref{pub} and Table~\ref{ind}, respectively.

Firstly, STAN achieves the most effective results on all tasks in both metrics and outperforms all the competitive baselines in the industrial and public datasets.
Secondly, the performance of $\mathrm{MLP_{Single}}$ lags behind all of the multi-task methods, indicating the benefit of joint optimization for multi-task learning.
Thirdly, by comparing different multi-task learning methods, we observe that the difficulty of optimizing different tasks can vary significantly.
The MMOE method only controls knowledge learned by shared-expert layers. Although it improves the $\mathrm{MLP_{Shared}}$ methods for some tasks, it suffers from a seesaw phenomenon \cite{mtl2020cgc} not only in different tasks but also in different metrics for the same task.
Thanks to the exploitation of specific-expert layers' learned knowledge, the vanilla PLE model outperforms the former models in most cases.
The performance of the AITM method is not consistent across all datasets, which may result from the characteristics of the dataset.
It can be seen that previous methods may decrease the NDCG score but increase the AUC score, which would require users to scroll down more times to discover their favorite items. This drawback may result from ignoring user stage information, which overlooks the real needs of the user.
Overall, our proposed STAN model achieves significant improvement compared to several state-of-the-art methods, demonstrating the efficiency of introducing adaptively-learned stage information into the recommendation system.

In particular, we evaluate the effect of the latent user stage representation module.
As shown in the result of \textit{STAN w/o Beta}, if we apply the predicted $\tilde{y}$ instead of the preference $\gamma$ to the overall loss update, the performance experiences greater fluctuation.
The reason could be the small number of action histories for the majority of users, as modeling the user's preference based solely on a few actions will lead to unstable results.

\subsection{In-depth Stage Analysis}

To answer \textbf{RQ2}, we conduct an in-depth analysis to uncover the real impact of incorporating \textit{user lifecycle} adaptive stage information into the recommendation process.

\subsubsection{Benefits of Considering Stages}

\begin{figure}[!htp]
\includegraphics[width=.9\columnwidth]{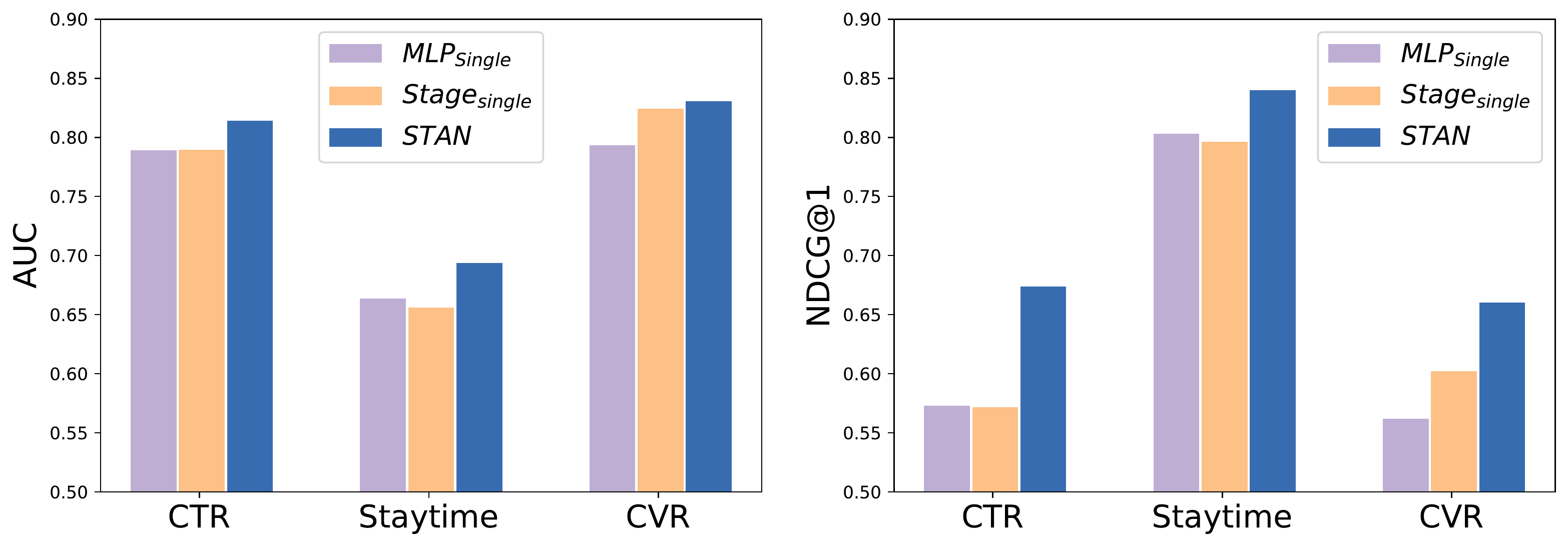}
\caption{Performance of predictions trained only on subsets separated by stage information for the industrial dataset.}
\label{stage_sub}
\end{figure}

To assess the benefits of incorporating stage information during training, we first divide the training dataset into three subsets based on user stage information and then perform separate prediction tasks on each of these subsets. We refer to this as $\mathrm{Stage_{Single}}$. The results are presented in Fig.~\ref{stage_sub}. Surprisingly, even though the subsets are significantly smaller than the original training dataset, the performance is nearly the same, and sometimes even better. In this case, each subset is approximately 1/3 of the training dataset.
This can be attributed to the fact that stage information helps reduce the number of noisy samples in the entire dataset. In multi-task learning problems, not every sample is useful for all tasks. However, conventional multi-task learning methods train all samples for each task simultaneously, which inevitably introduces noise for specific tasks. By leveraging stage information, multi-task learning methods can reduce the noise caused by instances from other stages during training. Consequently, they can learn more accurate representations, leading to improved performance.

\subsubsection{Benefits of Considering Adaptive Stages}

Although expert knowledge can be helpful in creating rule-based stage discrimination strategies, these designed rules may not be adaptive to various scenarios in applications, let alone different datasets. Therefore, we compare the performance of $\mathrm{PLE_{Stage}}$ and STAN in Table~\ref{pub} and Table~\ref{ind}, representing our framework under conditions of fixed user stage and adaptive user stage, respectively.
Generally, utilizing rule-based stage information results in inferior model performance compared to the model that employs adaptive stage information. This can be attributed to the differences in stage sensitivity between the models. As STAN represents the user stage using a learned vector instead of a predefined stage number, it can capture subtle variations in user stages more effectively.

\subsection{Case Study}

To address \textbf{RQ3}, we visualize the embedding vectors of users at different stages for both public and industrial datasets. 
For a clearer illustration, we create discrete stage labels by clustering the stage representation vectors.

\subsubsection{User Stage Visualization}

\begin{figure}[]
  \includegraphics[width=\columnwidth]{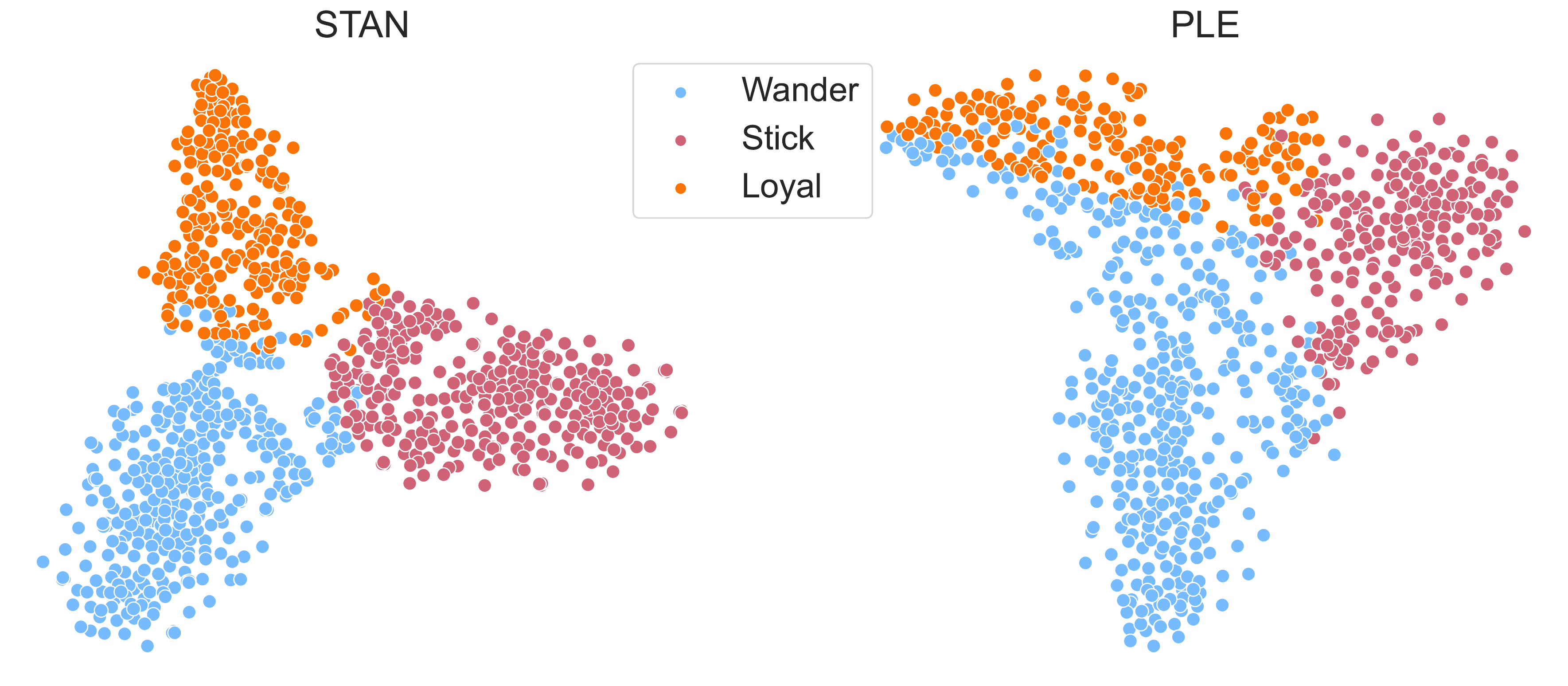}
  \caption{2D t-SNE\cite{van2008tsne} projections of the user embedding results of STAN and PLE on the public dataset. }
  \label{tsne_pub}
\end{figure}

For each dataset, we randomly select 1,000 user embeddings from the stages divided by the method in Sec.~\ref{data}.
Each point in Fig.~\ref{tsne_pub} represents a user with embeddings learned by STAN and PLE, respectively.
 Each color denotes a type of user stage; since the public dataset is pre-processed by its issuer, there are no users at Stage \textit{New}.
According to Fig.~\ref{tsne_pub}, the user embeddings learned by STAN can be more clearly separated from each other compared to those learned by PLE. This demonstrates the effectiveness of STAN in detecting user stages.

\subsubsection{User Stage in Change}

\begin{figure}[]
  \includegraphics[width=\columnwidth]{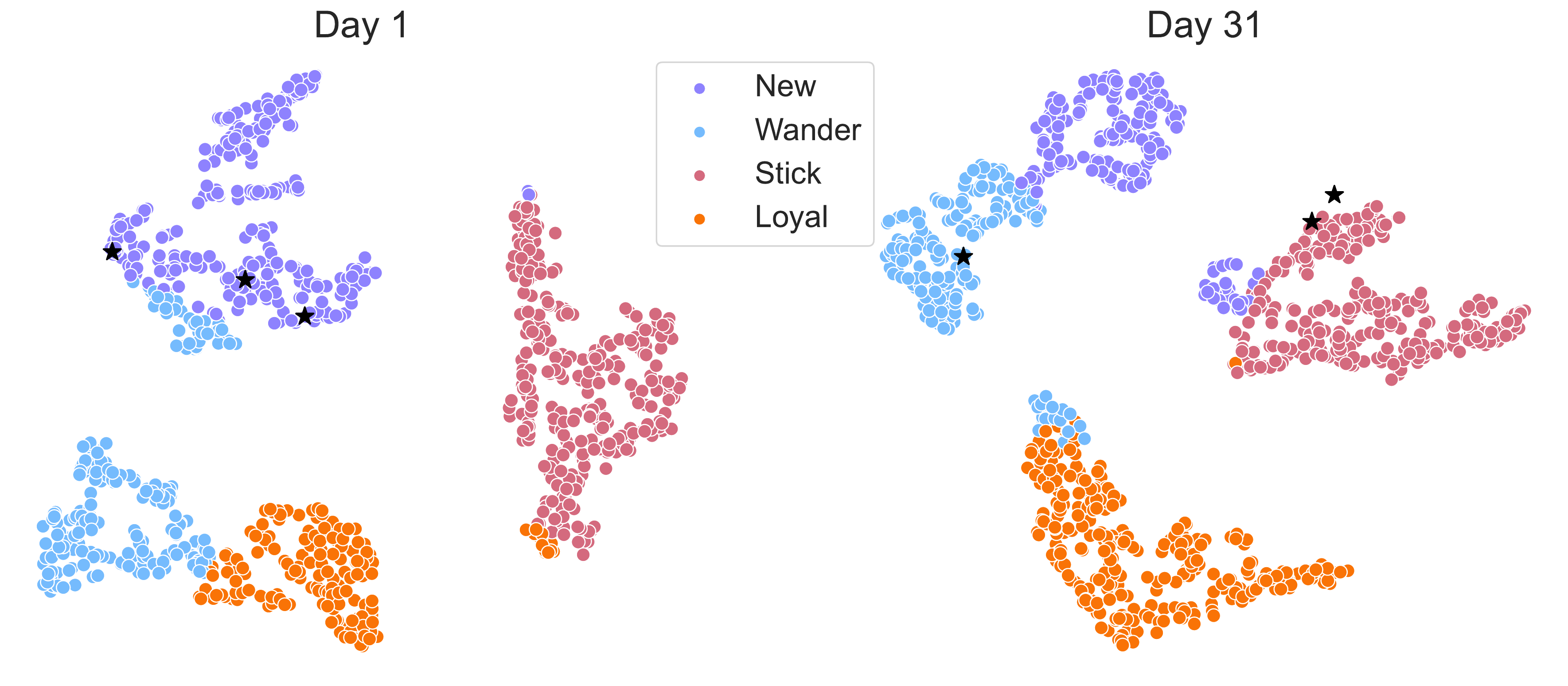}
  \caption{2D t-SNE\cite{van2008tsne} projection for visualizing user stage shifts over time on the industrial dataset. The black stars ($\bigstar$) denote the user from the same stage on Day 1.}
  \label{tsne_indus}
\end{figure}

A user's stage may change as he/she dwells on the platform. 
The \textit{user lifecycle} stage detection method should be adaptive to the shift of user stages as their features change. To evaluate this ability, we visualize the detected user stages of the same group of users on the first and last day of our industrial dataset in Fig.~\ref{tsne_indus}. The concerned users are highlighted as stars. We find that the same user may exhibit very different behaviors at different times, which indicates different user stages.
Take the users in Fig.~\ref{tsne_indus} as an example. At the start, all of them are at stage \textit{New}. However, one month later, one of them transitions to stage \textit{Wander}, and the other two transition to stage \textit{Stick}.
The observed user stage changes indeed highlight the dynamic nature of users' preferences and emphasize the importance of tracking these preferences in multi-task recommendation systems.

\subsection{Online A/B Testing and Deployment}
We carried out rigorous online A/B testing in our e-commerce live streaming scenario from 2023-03-15 to 2023-04-04, with a daily average of millions of users. Our proposed STAN model demonstrated significant improvements compared to its predecessor, with a CTR increase of 3.94\%, staytime increase of 3.05\%, and a CVR increase of 0.88\%. Based on these promising results, our model has been fully deployed and now serves the main traffic.

It is important to highlight that during the nearly month-long A/B testing, the e-commerce system underwent large-scale promotions, substantially impacting users' decision-making processes and potentially influencing their lifecycle stages. Despite these challenges, STAN's performance improvements remained consistent, further accentuating its effectiveness in managing user preferences in recommendation systems.

\section{Discussion}
\label{discussion}

In practical applications, it is a widespread industry approach to develop separate models for active and non-active users, often referred to as "cold-start users" \cite{recsys2022coldstart, son2016dealing}. The model for non-active users typically aims to boost engagement, while the model for active users concentrates on improving their CVR or Staytime. However, users in the same class can exhibit diverse characteristics. For instance, Alice and Bob, both active on the platform, might have significantly different CVRs, such as 0.7 and 0.2.

Thus, we propose a more nuanced approach to capture their preferences, better distinguishing between such users. This approach could benefit other recommendation structures, such as session-based user modeling techniques. Session-based recommendation approaches often assume that all labels are correct. However, since the user's preference is uncertain at some stages, user-generated labels at these stages should not be used as deterministic for modeling. In this scenario, incorporating our method could help session-based methods learn from more convincing labels, resulting in a more accurate representation of users' preferences and improved recommendations.

The former observation also leads us to another critical aspect largely overlooked in existing research: the trade-off between multiple task objectives. As these objectives are optimized globally for various goals, this approach can result in potential conflicts in gradient optimization. We argue that reducing the conflicts can be achieved by identifying the specific tasks that different users prioritize, allowing for a more tailored recommendation system that adapts to individual user preferences.

Building upon these insights, we have designed our experiments to capture the user's recent preference effectively. In our method's overall design, all samples preceding the current sample are used for calculating the average, which is applied to both public and industrial datasets. However, in a real online system, considering a too-long period would lead to higher overhead in storage computation, and behaviors with extended intervals may create perturbations to the current behavior and introduce instability, which is beyond the scope of this paper. Therefore, our online system focuses on behaviors within a 30-day window to calculate the average, ensuring a more accurate and manageable representation of users' recent preferences.

\section{Related Works}
\label{relatedWork}
% \cite{www2022contrastive}: CTR+readTime / CTR+CVR

In this section, we review related work in two main areas: (i) multi-task learning for recommendation systems and (ii) user representation modeling.

\subsection{Interested Tasks on Recommendation Systems}

Existing research on multi-task recommendation systems can be broadly classified into two categories. The first category focuses on user conversion rates \cite{sigir2018essm, sigir2020esm2, kdd2021aitm, sigir2022escm, www2022contrastive}, which yield significant profits for e-commerce platforms. The most representative user conversion tasks are the Click-Through Rate (CTR) and Click Conversion Rate (CVR). Some studies, such as \cite{sigir2018essm, sigir2022escm, www2022mlpr}, propose optimizing these tasks concurrently in a multi-task learning model, aiming to leverage valuable knowledge learned across tasks. However, the inherent divergence between prediction tasks can lead to reduced overall performance when optimizing them simultaneously. To address this issue, recent work \cite{wsdm2022leaving, www2022mlpr} has achieved notable success by incorporating scenario knowledge into task prediction. For example, Zhang et al. \cite{wsdm2022leaving} employed a meta-learning approach to predict tasks across multiple advertising scenarios. However, scenario information alone is insufficient for providing user-specific recommendations.

The second category of research aims to engage users on online platforms. Common metrics include user feedback, such as clicks, finishes, and dwell time \cite{recsys2014dwell, kdd2019reinforcement, mtl2020cgc, wu2021fairness}. Among these, user clicks and dwell time, or staytime, are the most descriptive metrics \cite{www2022feedrec}, but they are often overlooked in the context of the first category of work and rarely considered alongside CVR, which is essential for user conversion on online platforms.

\subsection{User Modeling and Representation in Lifecycle}

User modeling is another critical aspect of related work. It has received significant attention as user behavior variations offer valuable insights into user interests, particularly in the context of shifting trends. Some research models users based on their behavior sequences, using either Markov-chain methodologies \cite{www2010markov} or deep neural networks \cite{iclr2015rnn, wsdm2018cnn} to implicitly model user state dynamics and predict resulting behaviors. These methods primarily focus on short-term user modeling constrained by recent behaviors. To identify long-term behavior dependencies works such as \cite{ren2019lifelong, ali2020search, yuan2021one} have proposed recommendation system models with "lifelong" learning capacity. However, these models tend to learn long-term user behaviors coarsely, overlooking differences between various lifecycle stages.

Some studies have considered cold-start as a stage of the lifecycle for recommendation system users \cite{cao2022icde}. While these cold-start frameworks excel in serving new users, their performance diminishes as users mature. Unfortunately, model alteration often leads to user churn and instability in industrial applications.

\section{Conclusion}
\label{con}

% therefore: 前面的可以不看，只看这一句。一定要谨慎写
In this paper, we innovatively introduce the concept of \textit{user lifecycle} stages to enhance multi-task learning in recommendation systems. We propose STAN, a user lifecycle stage-adaptive framework that models latent stage information. STAN first learns user preferences for various tasks by utilizing user behaviors and then represents latent user stages based on these learned preferences. By incorporating latent user stage information into multi-task recommendations, STAN can identify the most critical task for each user and adjust accordingly as users' interests evolve. Experimental results on public and industrial datasets and online recommendation services demonstrate the effectiveness of our proposed framework.

For future work, we plan to explore the application of adaptive stage information in a broader range of contexts, including online social networks, advertising, and search systems.

% \section{Acknowledgments}

\bibliographystyle{ACM-Reference-Format}
\bibliography{sample-base}
\end{document}